\documentclass[journal=nalefd,manuscript=letter,layout=twocolumn]{achemso}

\usepackage[version=3]{mhchem} 
\usepackage{graphicx}
\usepackage{dcolumn}
\usepackage{bm}
\usepackage[colorlinks,linkcolor=blue,anchorcolor=blue,citecolor=blue,urlcolor=blue]{hyperref}
\usepackage{color,times}
\usepackage[mathlines]{lineno}
\usepackage{amssymb}
\usepackage{amsmath}
\usepackage{natbib}
\usepackage{soul,color}

\title{Unusually low thermal conductivity of
       atomically thin 2D tellurium}

\author{Zhibin Gao}
\affiliation{Center for Phononics and Thermal Energy Science,
             China-EU Joint Center for Nanophononics,
             Shanghai Key Laboratory of Special Artificial
             Microstructure Materials and Technology,
             School of Physics Sciences and Engineering,
             Tongji University, Shanghai 200092, China}

\author{Fang Tao}
\affiliation{Center for Phononics and Thermal Energy Science,
             China-EU Joint Center for Nanophononics,
             Shanghai Key Laboratory of Special Artificial
             Microstructure Materials and Technology,
             School of Physics Sciences and Engineering,
             Tongji University, Shanghai 200092, China}

\author{Jie Ren}
\email
            {Xonics@tongji.edu.cn}%
\affiliation{Center for Phononics and Thermal Energy Science,
             China-EU Joint Center for Nanophononics,
             Shanghai Key Laboratory of Special Artificial
             Microstructure Materials and Technology,
             School of Physics Sciences and Engineering,
             Tongji University, Shanghai 200092, China}

\date{\today} 


\begin{document}

\begin{abstract}
Tellurium is a high-performance thermoelectric material due to its
superior electronic transport and low lattice thermal conductivity
($\kappa_L$)~\cite{lin2016tellurium}.
Here, we report the ultralow $\kappa_L$ in the monolayer tellurium,
i.e., tellurene, which has been successfully synthesized in recent
experiments. %
We find tellurene has a compellingly low room temperature $\kappa_L$
of 2.16 and 4.08~W~m$^{-1}$~K$^{-1}$ along the armchair and
zigzag directions, respectively, %
which is lower than any reported values for other 2D
materials. We attribute this unusually low $\kappa_L$ to the soft acoustic
modes, extremely low-energy optical modes and the strong
scattering among optical-acoustic phonons, which place tellurene as
a potential novel thermoelectric material. Finally, we disclose
that $\kappa_L$ is proportional to the largest acoustic phonon
frequency ($\omega_{D}^{a}$) and the lowest optical phonon
frequency at $\Gamma$ point ($\omega_{\Gamma}^{o}$) in 2D
materials, which reflect both harmonic and anharmonic
thermal properties respectively.
\end{abstract}

\section{Introduction}

Graphene, maybe the most studied 2D system in
the history of science, displays record thermal
conductivity~\cite{balandin2008superior}, comparable to
single-wall carbon nanotubes~\cite{DT130}. Also other 2D materials
such as hexagonal BN display much higher thermal conductivities
than most 3D bulk materials. %
Here we show that atomically thin monolayer of Te, which has been
synthesized recently, has an unusually low lattice thermal
conductivity. Minimizing thermal conductivity is very important
for thermoelectrics to efficiently convert unavoidable waste heat
to electricity, since the figure of merit $zT$ is inversely
proportional to this quantity.

Specifically, the figure of merit of a thermoelectric material is
expressed as $zT ={S^2\sigma T/(\kappa_e+\kappa_L)}$, where $S,
\sigma, T, \kappa_e$ and $\kappa_L$ are the Seebeck coefficient,
electric conductivity, absolute temperature, electronic thermal
conductivity and lattice thermal conductivity, respectively.
Hunting for optimum \textit{zT} materials needs not only a maximum
power factor ($S^2\sigma$), but also a simultaneously minimum
thermal conductivity ($\kappa_e+\kappa_L$). Since electric
properties $S$, $\sigma$ and $\kappa_e$ couple strongly with each
other and interdepend in complicated ways, optimization of $zT$
becomes an arduous issue to realize the waste heat
recovery~\cite{snyder2008complex}. Fortunately, owing to the
different and separate scale of mean free paths of electrons and
phonons, $\kappa_L$ is a relatively independent parameter in $zT$.
Therefore, seeking materials of ultralow $\kappa_L$ becomes an
effective way to achieve high thermoelectric
performance~\cite{zhao2014ultralow}, and over the past decades,
considerable progress has been made in decreasing $\kappa_L$, to
realize so-called ``phonon-glass electron-crystal behavior".

As an accepted rule of
thumb~\cite{slack1973nonmetallic,lindsay2013first}, we sum up some of the
conditions in dielectric materials that can lead to ultralow $\kappa_L$:
(i) complex crystal structure (such as skutterudites~\cite{wei2017filling},
clathrates~\cite{zeier2016engineering,tadano2015impact}, embedded
nanoparticles~\cite{zhao2017superparamagnetic}), (ii) large
average atomic mass, (iii) weak interatomic bonding, and (iv)
strong anharmonicity (such as SnSe~\cite{li2015orbitally}). A
small Debye temperature, $\theta_D$, always originates from a
combination of heavy elements (ii) and low atomic coordination
(iii)~\cite{carrete2014low}. Furthermore, ultralow
$\kappa_L$ can be also obtained through phonon-liquid in
copper ion~\cite{liu2012copper}, resonant bonding (such as
rocksalt group IV-VI compounds~\cite{lee2014resonant} and
in-filled CoSb$_3$~\cite{zhao2015multi}) and lone electron pairs
(such as group I-V-VI$_2$ compounds~\cite{nielsen2013lone} and
InTe~\cite{jana2016origin}).


On one hand, bulk Te has been recently shown as a superior
thermoelectric material with $zT=1.0$~\cite{lin2016tellurium} in
addition to as a topological
insulator~\cite{agapito2013novel,hirayama2015weyl}, since the room
temperature $\kappa_L$ %
is experimentally measured as low as $1.96\scriptsize{\sim}
3.37$~W~m$^{-1}$~K$^{-1}$~\cite{ho1972thermal}. Recently,
Zhu \textit{et al.}~\cite{zhu2017multivalency}
and Chen \textit{et al.}~\cite{chen2017ultrathin}
have successfully synthesized ultrathin layers tetragonal
$\beta$-tellurene on highly oriented pyrolytic graphite (HOPG)
by using molecular beam epitaxy, which has much larger carrier mobility
than MoS$_2$ and is highlighted in an exclusive report for its
potential implications~\cite{naturenews}.
Moreover, Liu \textit{et al.}~\cite{liu2017tellurium}
and Qiao \textit{et al.}~\cite{qiao2018few} also
indicate few layer tellurene has extraordinarily electronic
transport properties and can be made high-performance field-effect
transistors by Wang \textit{et al.}~\cite{wang2017large}.
Yet, $\kappa_L$ of this intrinsic 2D
tellurene structure is far from clear. In this letter, we
explore intrinsic $\kappa_L$ of monolayer $\beta$-tellurene.

On the other hand, Dresselhaus \textit{et al.} have pointed out
the low dimensional materials (such as 2D materials) can further
enhance the electronic performance comparing to the 3D
counterparts due to quantum confinement~\cite{dresselhaus2007new}.
However,  2D materials usually also have larger $\kappa_L$ than
their 3D bulk counterparts due to significant contributions of
out-of plane (ZA) modes~\cite{seol2010two,balandin2011thermal}.
This may counteract
the huge potential of a 2D material with better thermoelectric
\textit{zT}. As such, seeking 2D materials of ultralow $\kappa_L$ is
significant to fabricate superior and miniaturized thermoelectric
devices~\cite{Franklin}. Therefore, although 2D tellurene is
supposed to have good electronic properties, its thermal
properties are crucial and will strongly impact tellurene's
potential to possess good thermoelectric performance.

In this Letter, we find that tellurene has unusually low room
temperature $\kappa_L$ that of merely 2.16 and 4.08~W~m$^{-1}$~K$^{-1}$
along the armchair and zigzag directions, respectively.
Those values, although obtained from the 2D crystalline tellurene,
are comparable to the bulk Te, which is quite
counterintuitive given the well-known trend that 2D materials have
usually larger $\kappa_L$ than their 3D counterpart due to the
significant ZA mode contribution to the $\kappa_L$
~\cite{seol2010two,balandin2011thermal}. Moreover, we find
tellurene has the lowest
recorded $\kappa_L$ among the 2D materials family to date.
Therefore, we carefully scrutinize the underlying mechanism of
ultralow $\kappa_L$ of tellurene from the aspects of
harmonic and anharmonic properties in the following.


\begin{figure}
\includegraphics[width=0.9\columnwidth]{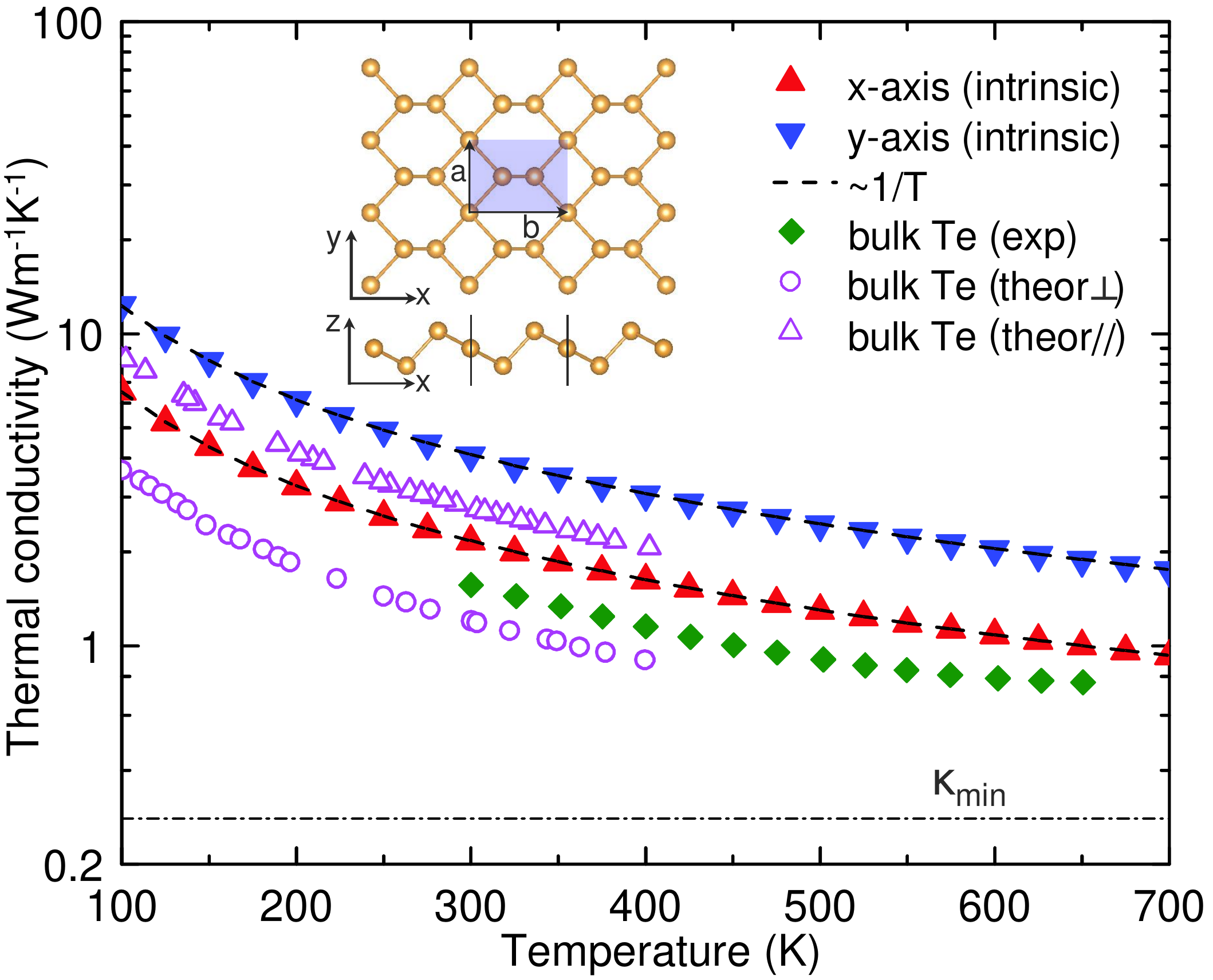}
\vspace{-2mm} %
\caption{Lattice thermal conductivity of tellurene
as a function of temperature. Ball and stick model of the
tellurene in top and side views are shown in the inset. The
primitive cell is indicated by the blue shading in the top view.
$\bm{a}$ and $\bm{b}$ are the lattice vectors spanning the 2D
lattice.  Black dashed lines are~1/\textit{T} fitting of
temperature dependent $\kappa_L$. Green rhombic dots and purple
triangles/circles are the experimental~\cite{lin2016tellurium} and
theoretical~\cite{peng2015anisotropic} (parallel/perpendicular to
the bulk helical chains) data of bulk Te, and dashed lines are
provided as a guide to the eye. Red and blue solid triangle
are the intrinsic $\kappa_L$ we obtained
from phonon Boltzmann transport equation considering phonon-phonon
scattering. The dash dotted line is the lower limit $\kappa_{min}$
of bulk Te according to the Cahill
model~\cite{cahill1992lower,lin2016tellurium}. \label{fig1}}
\end{figure}

\begin{figure}
\includegraphics[width=1.0\columnwidth]{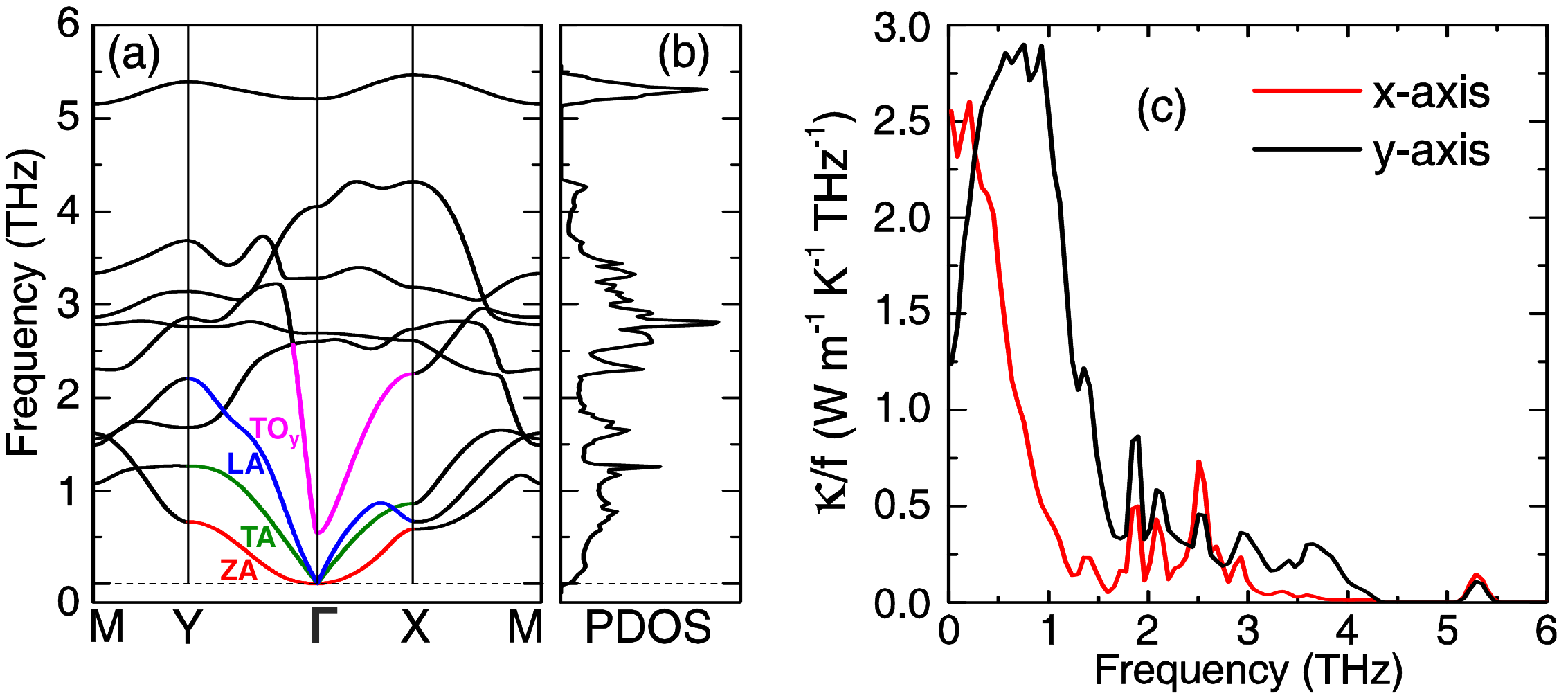}
\vspace{-8mm} %
\caption{(a) Phonon band structures, (b) Phonon density of states
(PDOS) of tellurene. Three acoustic phonon branches, which
originate from the $\Gamma$-point of the Brillouin zone,
correspond to an out-of plane (ZA) mode, an in-plane transverse
(TA) mode, and in-plane longitudinal (LA) mode. Asymmetric
waterfall-like transverse optical phonon mode along y axis
(TO$_y$) is also marked. The dashed lines are provided as
a guide to the eye. (c) Frequency-resolved thermal conductivity
for tellurene in x and y directions at room temperature. \label{fig2}}
\end{figure}

\section{Lattice thermal conductivity}

In semiconductor and insulator, heat is mainly carried by phonons.
The anisotropic in-plane lattice thermal conductivity under the
relaxation time approximation can be calculated as sum of
contribution of all phonon mode $\lambda$ with wave vector \textbf{q}:
\begin{equation} %
\label{kappa}
\kappa_{\alpha\beta} = \frac{1} {V} \sum_{\lambda} C_\lambda
                       \upsilon_{\lambda \alpha}
                       \upsilon_{\lambda \beta} \tau_{\lambda},  \\
\end{equation}
where \textit{V} is the crystal volume, \textit{C$_\lambda$}
is the specific heat per mode, $\upsilon$$_{\lambda \alpha}$
and $\tau$$_{\lambda}$ are the velocity component along
$\alpha$ direction and the phonon relaxation time.
$\kappa_L$ can be obtained by solving the phonon Boltzmann
transport equation that is related to the harmonic and anharmonic
interatomic force constants.

$\kappa_L$ is an intensive property. Hence, a value
of thickness needs to be chosen in 2D materials when comparing
with the 3D counterpart. In order to make it more clear and
consistent, we also use the thermal sheet conductance (``2D
thermal conductivity'') with unit~W~K$^{-1}$ as that is the most
unequivocal variable in 2D materials. The thickness of
tellurene (6.16~{\AA}) is taken as the summation of the buckling
distance and the van der Waals (vdW) radii of Te
atom~\cite{gao2017novel,wu2017characterize}
that is in good agreement with our calculated value obtained by
artificially stacking tellurene layers.

Figure~\ref{fig1} shows the calculated $\kappa_L$ of tellurene along
armchair (x-axis) and zigzag (y-axis) directions as a function of
temperature, as well as the collected $\kappa_L$ data of bulk Te for
comparison. The intrinsic $\kappa_L$ (only consider the
phonon-phonon scattering) at room temperature of tellurene along x
and y directions are 2.16 and 4.08~W~m$^{-1}$~K$^{-1}$, respectively. As we
mentioned above, due to the symmetry $\kappa_L$ of bulk Te is
isotropic parallel to the helical chains but anisotropic when
perpendicular to the
chains~\cite{peng2015anisotropic,peng2014elemental}. We compare
the $\kappa_L$ between tellurene and bulk Te parallel to the helical
chains. These values of tellurene (red solid
triangle) are comparable to and even smaller than the
theoretical result ~\cite{peng2015anisotropic} of bulk Te along
the helical chains direction (open purple triangle) in all
temperature range, which is in contrast to what happens in
other layered 2D materials (such as graphene comparing to
graphite). Furthermore,
$\kappa_L$ along x direction is only one half of that in y
direction, indicating a large anisotropic thermal transport in
tellurene.

The minimum lattice thermal conductivity $\kappa_{min}$
of bulk Te according to the Cahill
model~\cite{cahill1992lower,lin2016tellurium} is shown as
reference. Additionally, in real experiment and practical devices,
boundary scattering is an important factor to the $\kappa_L$ of a
material with finite size (discussed in the Supporting
Information). Ultralow $\kappa_L$ in tellurene comparable with bulk
Te deviates from the well-known trend that 2D materials have
usually larger $\kappa_L$ than their 3D
counterpart~\cite{seol2010two,balandin2011thermal}, encouraging us
to explore the physical reason behind it.

The thermal sheet conductance of tellurene along x and y directions
are 1.33 and 2.51~nW~K$^{-1}$ which are also the lowest values in 2D
crystalline family to date (Supporting Information).
Furthermore, we find $\kappa_L$ of tellurene follows well with
\textit{T$^{-1}$} behavior, indicating a dominant Umklapp process
of phonon scattering that causes thermal resistivity. This nice
\textit{T$^{-1}$} curve is common in other heavy
elements~\cite{goldsmid2013thermoelectric} and recently is also
experimentally observed in bulk Te~\cite{lin2016tellurium}. The
unusually low and anisotropic $\kappa_L$ of tellurene will be
explained physically from both the harmonic and anharmonic
properties in following sections.

\begin{figure*}
\includegraphics[width=1.8\columnwidth]{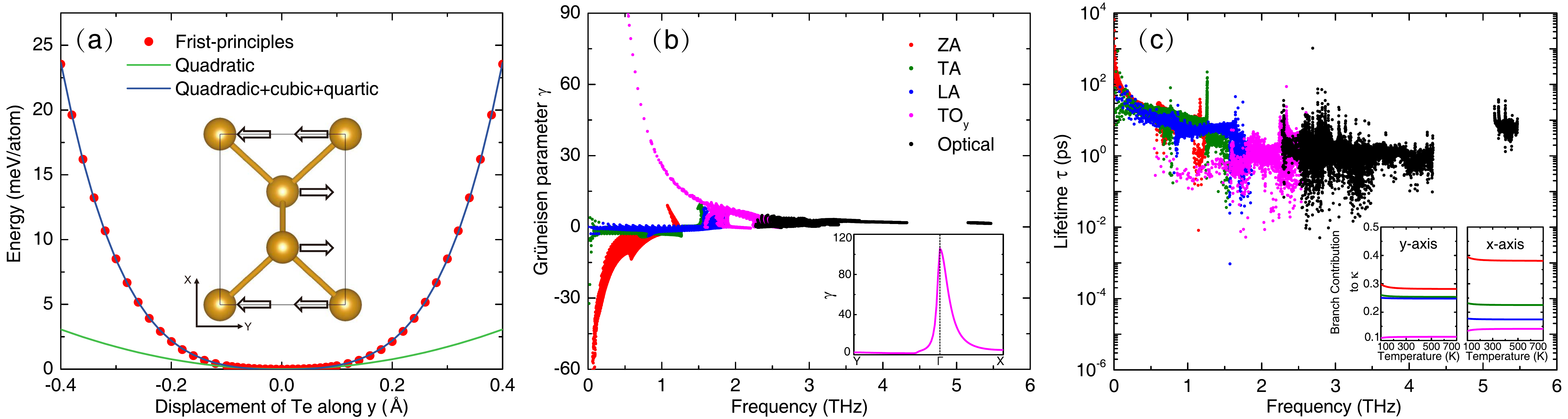}
\vspace{-2mm} \caption{(a) Anharmonic frozen-phonon potential with
quadratic and a polynomial fitting. The inset shows the vibration
direction of lowest-energy TO$_y$ phonon mode. (b) Gr\"{u}neisen
parameter $\gamma$ as a function of the phonon frequency. The
inset shows the $\gamma$ of waterfall-like optical phonon mode
(TO$_y$) along both directions. (c) Dependence of phonon
relaxation time on frequency at room temperature. The inset shows
the normalized contribution of each acoustic and TO$_y$ phonon
modes to the $\kappa_L$ as a function of temperature along both
directions.\label{fig3}}
\end{figure*}

\section{Soft Harmonic properties}

Tellurene has three atoms in each unit cell as shown in the inset
of Fig.~\ref{fig1}, so possesses three acoustic and six
optical phonon modes. For 2D materials, in the long wavelength
limit, very close to the $\Gamma$ point, the LA and TA modes are
linear in \textbf{q}, whereas the ZA mode is
quadratic~\cite{carrete2016physically}, with coefficients given by
2D continuum elasticity theory~\cite{DT255}. But when \textbf{q}
is slightly far away from $\Gamma$, ZA mode will have a
near-linear trend. As a matter of fact, calculated ``raw''
dynamical matrix often give an imperfect parabolic ZA mode and
sometimes even small artificially imaginary frequencies around the
$\Gamma$ due to the insufficient accuracy (supercell and k meshes)
in the simulation. Hence, we corrected force constants to
rigorously apply the translation and rotation
symmetries~\cite{carrete2016physically}. We calculated the
flexural rigidity \textit{D}($\Gamma$-X) and
\textit{D}($\Gamma$-Y) of~0.37 and~0.40 eV in tellurene along x
and y directions, which describes the flexural response to
out-of-plane stress of materials (Supporting Information).
Those values of tellurene are
about a quarter of graphene (1.4~eV~\cite{DT255}) and phosphorene
(1.55 eV~\cite{DT255}), indicating tellurene is much softer than
graphene and phosphorene.

Phonon dispersion and phonon density of states (PDOS) of tellurene
are shown in Fig.~\ref{fig2}a and Fig.~\ref{fig2}b. One can see
that, LA and TA phonon modes along $\Gamma$-X are much lower than
in $\Gamma$-Y direction, implying a smaller $\theta_D$ in x
direction. Moreover, an asymmetric optical phonon branch in
magenta line, like a waterfall, suddenly falls into the very low
frequency region. We find the corresponding optical mode vibrates
along y direction as shown in the inset of Fig.~\ref{fig3}a, so
that we call it TO$_y$ in the following. Group velocities, defined
as $\upsilon = {\partial \omega}/{\partial q}$ are shown in the
Supporting Information. This ultra-small sound velocities will
contribute to the reason that tellurene has unusually low $\kappa_L$
because $\kappa_L$ is proportional to the $\upsilon^2$ based on
Eq.~\eqref{kappa}.

From the tellurene's phonon spectrum, the ZA mode along both
directions is much flat (smaller \textit{D})~\cite{DT255},
possessing much lower $\upsilon$. As we all know, ZA mode plays a
crucial role in the $\kappa_L$ of 2D
materials~\cite{seol2010two,balandin2011thermal}. For instance,
75\% $\kappa_L$ derived from the ZA mode in
graphene~\cite{lindsay2010flexural}. Thus, abundant such soft ZA
modes with much lower $\upsilon$ significantly weaken its role in
the thermal conductivity of tellurene, which is another cause for
ultralow $\kappa_L$. We also calculated the frequency-resolved
$\kappa_L$ for tellurene. Similar to graphene and other 2D
materials, in tellurene low frequency phonons dominate the
contribution of $\kappa_L$ in both x and y directions shown in
Fig.~\ref{fig2}c. As we will discussed later, waterfall-like
TO$_y$ mode enhance the scattering between acoustic and optical
phonon modes in tellurene. Therefore, the contribution of the
acoustic phonon modes to the total $\kappa_L$ will be weaken.

The mechanical properties of tellurene calculated based on elastic
solid theory~\cite{gao2017novel} are shown in Supporting Information.
Young's modulus and Poisson's ratio of tellurene in y direction is
about two times larger than that in x direction.
Tellurene has very small \textit{E} and Poisson's ratio $\nu$ along both
directions, indicating a lower vibrational
strength~\cite{xiao2016origin}. As we discussed above, a small
$\theta_D$, means average low phonon frequency.
A small $\theta_D$, combined with low
$\upsilon$, always implies a weak interatomic bonding, which will
decrease the $\kappa_L$ of heavy tellurene (criterion ii and iii).
This is another causation from harmonic properties that why
tellurene has an unusually low and anisotropic $\kappa_L$.

\section{Giant Gr\"{u}neisen parameter}

Strong anharmonicity in materials can lead to low $\kappa_L$
(criterion iv). The Gr\"{u}neisen parameter, $\gamma$, measures the
effect of volume changing of a crystal upon the thermal expanded
phonon vibrations so that large $\gamma$ indicates a large bonding
anharmonicity in materials. This giant anharmonicity of TO$_y$
phonon mode is strongly related to the chemical bonding and
distortion potential~\cite{li2015orbitally} shown in the
Fig.~\ref{fig3}a. We move all atoms along its eigenvector and the
symmetric potential intensely deviates from the quadratic
function. The strong anharmonicity of TO$_y$ phonon mode can be
further confirmed by a polynomial fitting, which is consistent
with the giant $\gamma$ in the inset of Fig.~\ref{fig3}b. The
large $\gamma$ of TO$_y$ phonon branch enhances the scattering
rates and anharmonicity, leading to the ultralow $\kappa_L$ of
tellurene.

To further explore the $\gamma$ distribution, we calculated
$\gamma$ for whole frequency spectrum in Fig.~\ref{fig3}b. Below
the frequencies of~0.5 THz (see also in Fig.~\ref{fig2}a), there
is no obvious acoustic-optical (a-o) phonon scattering and ZA mode
has the largest $\gamma$. As a matter of fact, $\kappa_L$ is both
proportional to the anharmonic interactions (matrix) elements
and the inverse of phase space volume \emph{P$_3$}. The former
is closely related to the frequency-dependent $\gamma$ and the
latter describes all available three-phonon scattering
processes that need to satisfy the energy and momentum conservation
simultaneously~\cite{lindsay2008three,lee2014resonant,ShengBTE2014}.
The calculated \emph{P$_3$},
shown in the Supporting Information, indicates
that three-phonon scattering channels in tellurene does not vary
too much. Hence, ultralow $\kappa_L$ mainly stems from the large
$\gamma$, rather than the increase of scattering channels.



In the frequency range of 0.5$\scriptsize{\sim}$2.0~THz, $\gamma$
of ZA, TA, LA and TO$_y$ phonon modes suddenly jump, indicating a
giant anharmonic scattering change of interactions (matrix)
elements between the acoustic and optical phonons.
This phenomenon can also be mapped in the phonon dispersion shown
in Fig.~\ref{fig2}a. The giant $\gamma$ induced by the large a-o
phonon scattering is also the origin of ultralow $\kappa_L$ of in
SnSe~\cite{li2015orbitally}, single-layer transition metal
dichalcogenides~\cite{gu2014phonon},
phosphorous~\cite{qin2016resonant}, rocksalt
structure~\cite{lee2014resonant}, which when in absence leads to
the ultrahigh $\kappa_L$ of boron arsenide~\cite{lindsay2013first}.

In the inset of Fig.~\ref{fig3}b, we plot the corresponding
$\gamma$ for TO$_y$ phonon mode along x and y directions. One can
see clearly an asymmetric giant $\gamma$ along both directions and
$\gamma$ in x direction is relatively larger than that in y
direction, indicating a stronger optical-acoustic phonon
scattering so that a lower $\kappa_L$ in x direction in the
frequency range of 0.5$\scriptsize{\sim}$2.0~THz [see
Fig.~\ref{fig2}(c)]. Therefore, in x direction the stronger
anharmonic scattering together with the weaker harmonic properties
both lead to the unusually low $\kappa_L$ in x direction, resulting
an anisotropic $\kappa_L$ in tellurene.

\section{Strong optical-acoustic phonon scattering}

A finite $\kappa_L$ is an outcome of the phonon-phonon
scattering~\cite{gao2016heat,gao2016stretch}. As evident in
Fig.~\ref{fig3}c, in the range of 0.5$\scriptsize{\sim}$2.0~THz,
the calculated phonon lifetimes of ZA, TA, LA and TO$_y$ phonon
modes are significantly shortened due to the strong a-o scattering. %
A smaller phonon lifetime will result in a smaller $\kappa_L$
according to Eq.~\eqref{kappa}, and the contribution of these four
phonon branches is shown in the inset of Fig.~\ref{fig3}c. When
putting graphene on substrate, the room temperature
$\kappa_L$ will significantly decrease from
$3000\scriptsize{\sim}5000$~W~m$^{-1}$~K$^{-1}$ of suspended
graphene~\cite{balandin2008superior} to 600~W~m$^{-1}$~K$^{-1}$ of
supported graphene~\cite{seol2010two,balandin2011thermal}, due to
the large suppression of the ZA mode contribution by
substrates~\cite{seol2010two}, since 75\% of graphene's $\kappa_L$
are carried by ZA phonon mode~\cite{lindsay2010flexural}. Due to
the strong a-o scattering in tellurene, ZA mode contribution has
been reduced to the 38.2\% and 28.3\% along x and y directions.
For tellurene, the ZA mode contribution is largely suppressed by
the soft dispersion, along with the large scattering from TO$_y$
phonon mode, finally resulting in the ultralow $\kappa_L$. When
considering tellurene of finite size, the boundary scattering will
further decrease the $\kappa_L$. The corresponding
size-dependent $\kappa_L$ calculation for suspended
tellurene is shown in the Supporting Information.

\begin{figure}
\includegraphics[width=1.0\columnwidth]{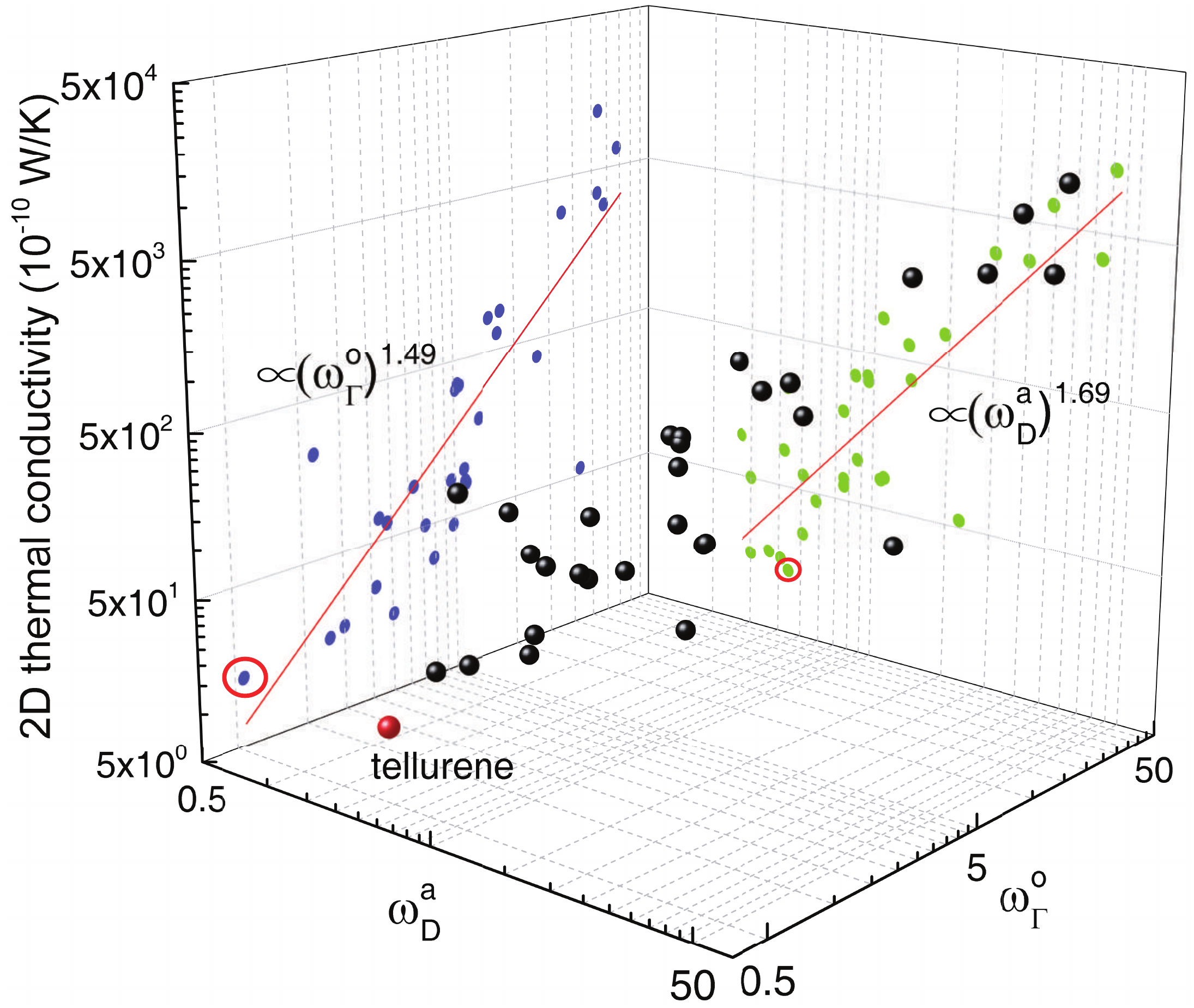}
\vspace{-7mm} %
\caption{Room temperature 2D $\kappa_L$ in W/K as a function of
$\omega_{D}^{a}$ and $\omega_{\Gamma}^{o}$ that are the largest
acoustic phonon frequency and the lowest optical phonon frequency
at $\Gamma$ point in THz. The projected data are fitted as
$\kappa_L$ $\propto$ ($\omega_{D}^{a}$)$^{1.69}$ and $\kappa_L$
$\propto$ ($\omega_{\Gamma}^{o}$)$^{1.49}$. Tellurene is marked in
red with the lowest $\kappa_L$. \label{fig4}}
\end{figure}

Moreover, Fig.~\ref{fig4} shows tellurene has the lowest $\kappa_L$
based on our collected data (Supporting Information) of 2D materials.
We find that the largest acoustic phonon frequency
($\omega_{D}^{a}$) and the lowest optical phonon frequency at
$\Gamma$ point ($\omega_{\Gamma}^{o}$) are two good
descriptors to estimate $\kappa_L$.
On the one hand, $\omega_{D}^{a}$ reflects linear part of
thermal transport (harmonic approximation). A lower
$\omega_{D}^{a}$ means a relatively lower group
velocities and softer acoustic phonon
vibrations~\cite{zhao2014ultralow}, which will lead to a
lower $\kappa_L$ based on Eq.~\eqref{kappa}.
On the other hand, $\omega_{\Gamma}^{o}$ indicates the
gap between the acoustic and lowest optical phonons.
Optical phonons provide scattering channels
for the acoustic branches~\cite{lindsay2013first} and
have a effect on the anharmonic interactions (matrix)
elements.
Hence, $\omega_{D}^{a}$ and $\omega_{\Gamma}^{o}$
are
two good characteristics of $\kappa_L$ from both harmonic
and anharmonic aspects. After projection, we find 2D
$\kappa_L$ can be fitted as
$\kappa_L$ $\propto$ ($\omega$$_{D}^{a})^{1.69}$ and
$\kappa_L$ $\propto$ ($\omega$$_{\Gamma}^{o})^{1.49}$.
Note that two exponents in the above trends exist
some uncertainty based on the limitedly published
values of $\kappa_L$, thus 1.69 and 1.49 are not
very rigorous but the trends between $\kappa_L$
and $\omega_{D}^{a}$ and $\omega_{\Gamma}^{o}$
are universal (at least, positive correlations).

These two trends are different from the model
$\kappa_L$ $\propto$ ($\omega_{D}^{a}$)$^3$ described
by Slack \textit{et
al.}~~\cite{slack1973nonmetallic,goldsmid1965thermal}
who proposed that $\kappa_L$ of bulk materials
above the Debye temperature and governed by the Umklapp
phonon scattering can be written in terms of $\omega_{D}^{a}$:
\begin{equation}%
\label{model}
\kappa_L \propto  \frac{a^4 \rho (\omega_{D}^{a})^3} {\gamma^2 T},  \\
\end{equation}
where \textit{a}$^3$, $\rho$ and $\gamma$ are the average volume
occupied by one atom of the crystal, density and the acoustic
phonon Gr\"{u}neisen parameter. While in 2D materials, our finding
shows $\kappa_L$ $\propto$ ($\omega_{D}^{a}$)$^{1.69}$ that violates
the above Slack formula. Hence it
should be careful when applying Slack model to $\kappa_L$ of 2D
materials~\cite{ma2014examining}. In addition, note that
$\gamma$ is also a function of $\omega_{D}^{a}$ and there
is no explicit relation with $\omega_{D}^{a}$, and has
also strong material dependence~\cite{clarke2003materials}.

From the phonon dispersion, PDOS of linear acoustic modes in bulk
materials is proportional to ($\omega_{D}^{a}$)$^2$, while in 2D
materials PDOS of quadratic ZA mode is a constant. In the
transition from 3D to 2D materials, there exists a transformation
of ZA mode from a linear dispersion in 3D to a quadratic one in 2D
materials. %
Based on the relation: $\kappa_L \propto \int_{0}^{\omega_{D}^{a}}
g(\omega) \upsilon^2 \tau d\omega$, in which g($\omega$) is PDOS,
we can find PDOS of 2D materials is a superposition of two linear
LA, TA and one unusual parabolic ZA phonon modes. Obviously,
constant PDOS of ZA mode will weaken the exponent between $\kappa_L$
and $\omega_{D}^{a}$ according to the Slack Eq.~\eqref{model}.

Nevertheless, these two trends reveal that
$\omega_{D}^{a}$ and $\omega_{\Gamma}^{o}$ are two relevant
descriptors for $\kappa_L$. $\omega_{D}^{a}$ reflects the strength
of acoustic phonon vibrations and group velocities.
$\omega_{\Gamma}^{o}$ discloses the important gap between acoustic
and optical modes that is very crucial for the optical-acoustic
phonon scattering rates and scattering
channels~\cite{ShengBTE2014,lindsay2013first,lee2014resonant}.
A lower $\omega_{\Gamma}^{o}$ will enhance the three-phonon
scattering processes and will have a significant impact on
the anharmonicity and attenuation on $\kappa_L$.

\section{Conclusion}

In conclusion, we have theoretically explored the unusually low
thermal properties of tellurene by the first-principle
calculations and phonon Boltzmann transport. To trace the ultralow
$\kappa_L$, we unveil the reasons from both the harmonic and
anharmonic aspects. Tellurene consists of heavy atomic mass
(criterion ii). From the harmonic view of phonon dispersion and
elasticity, low Debye temperature, group velocities of acoustic
phonons, Young's modulus, and shear modulus reveal the weak phonon
vibrations and interatomic bonding that lead to the unusually low
$\kappa_L$ in tellurene (criteria iii). For anharmonicity, large
$\gamma$, strong acoustic-optical phonon scattering and large
phonon-phonon anharmonic scattering rates are shown to illustrate
the strong anharmonicity in tellurene (criterion iv). These
convincing evidence has verified the unusually low $\kappa_L$ in
atomically thin 2D tellurium. Finally, we find $\kappa_L$ is
proportional to the largest acoustic phonon
frequency ($\omega_{D}^{a}$) and the lowest optical phonon
frequency at $\Gamma$ point ($\omega_{\Gamma}^{o}$) for
reported 2D materials. These two frequencies reflect
the thermal properties from both harmonic and anharmonic aspects,

Coupled with the superior electronic
transport~\cite{zhu2017multivalency,lin2016tellurium,wang2017large,liu2017tellurium,qiao2018few},
we hope ultralow $\kappa_L$ tellurene would shed a light on the implication for
thermoelectric field in the future.
Thickness-dependent $\kappa_L$ is also be an
interesting open question to understand the thermal
transport property in few-layer tellurene.

\section{Methods}

Our quantitative predictions are obtained by performing the
density functional theory (DFT) and by solving the phonon
Boltzmann transport equation. We performed density functional
theory calculations as implemented in the Vienna Ab initio
simulation package (VASP)~\cite{VASP1,VASP2} with a plane-wave
cutoff of 300 eV, 70\% higher than the maximum recommended cutoff
for the pseudopotentials. Perdew-Burke-Ernzerhof (PBE)
exchange-correlation functional \cite{PBE} along with the
projector-augmented wave (PAW) potentials \cite{PAW1,PAW2} are
used. Energy convergence value in self-consistent field (scf) loop
is selected as 10$^{-8}$ eV and a maximum Hellmann-Feynman forces
is less than 0.001~meV/{\AA}. Harmonic interatomic force constants
(IFCs) is obtained using Phonopy~\cite{phonopy} with
10$\times$10$\times$1 supercell, while
ShengBTE~\cite{ShengBTE2014,thirdorder2012,Gaussian2012} is
utilized to extract the anharmonic IFCs by solving the linearied
phonon Boltzmann transport equation. Converged cutoff of 0.55 nm
for the interaction range and q-grid of 100$\times$100$\times$1
are employed after testing. A 4$\times$4$\times$1 supercell with
3$\times$3$\times$1 Monkhorst-Pack k-point mesh is used for IFCs
calculations. For the correction of IFCs, we enforce the
translation and rotation symmetries to obtain a parabolic
out-of-plane (ZA)
mode~\cite{carrete2016physically} (Supporting Information).

\begin{suppinfo}
\end{suppinfo}
\quad\par

{\noindent\bf Author Information}\\

{\noindent\bf Corresponding Author}\\
$^*$E-mail: {\tt Xonics@tongji.edu.cn}

{\noindent\bf ORCID}\\
Zhibin Gao: 0000-0002-6843-381X \\
Jie Ren: 0000-0003-2806-7226

{\noindent\bf Notes}\\
The authors declare no competing financial interest.


\begin{acknowledgement}
Z.G. is grateful for the hospitality of Prof. David Tom\'{a}nek,
Michigan State University, where this work was initiated.
We thank David Tom\'{a}nek, Jes\'{u}s Carrete and Dan Liu
for very helpful discussions and their critical reading of
the manuscript.
Z.G., F.T. and J.R. were supported by the National
Natural Science Foundation of China with grant No.
11775159, the National Youth 1000 Talents Program in China, and
the startup Grant at Tongji University.
Computational resources have been provided by the Tongji
University and  Michigan State University
High Performance Computing Center. Z.G. gratefully acknowledges the China
Scholarship Council (CSC) for financial support (to C.S.,
201706260027).
\end{acknowledgement}

\providecommand{\latin}[1]{#1}
\makeatletter
\providecommand{\doi}
  {\begingroup\let\do\@makeother\dospecials
  \catcode`\{=1 \catcode`\}=2 \doi@aux}
\providecommand{\doi@aux}[1]{\endgroup\texttt{#1}}
\makeatother
\providecommand*\mcitethebibliography{\thebibliography}
\csname @ifundefined\endcsname{endmcitethebibliography}
  {\let\endmcitethebibliography\endthebibliography}{}

\end{document}